# Overlooked centrifugal and Coriolis forces in the atmosphere due to the Earth's motion in the solar system

Joseph Egger[1], Meteorological Institute, University Munich, Germany

Klaus Fraedrich[2], Max Planck Institute of Meteorology, Hamburg, Germany

**Abstract**: The slow revolution of the Earth and Moon around their barycentrum does not induce Coriolis accelerations. On the other hand, the motion of Sun and Earth is a rotation with Coriolis forces which appear not to have been calculated yet, nor have the inertial accelerations within the system of motion of all three celestial bodies. It is the purpose of this contribution to evaluate the related Coriolis and centrifugal terms and to compare them to the available atmospheric standard terms. It is a main result that the revolution is of central importance in the combined dynamics of Earth, Moon and Sun. Covariant flow equations are well known tools for dealing with such complicated flow settings. They are used here to quantify the effects of the Earth's revolution around the Earth-Moon barycenter and its rotation around the Sun on the atmospheric circulation. It is found that the motion around the Sun adds time dependent terms to the standard Coriolis forces. The related centrifugal accelerations are presented. A major part of these accelerations is balanced by the gravitational attraction by Moon and Sun, but important unbalanced contributions remain. New light on the consequences of the Earth's revolution is shed by repeating the calculations for a rotating Earth-Moon pair. It is found that the revolution complicates the atmospheric dynamics.

---

[1] Veronica.Egger@biologie.uni-regensburg.de

[2] Klaus.Fraedrich@mpimet.mpg.de

# 1. Introduction

Sun, Earth and Moon constitute a three-component system of central importance for atmospheric motion. As is well known, the Earth is spinning around its axis, revolving around the Earth-Moon barycenter and rotating around the Sun (Fig. 1). This motion of the Earth in the solar system is described by a fairly complicated guiding velocity. In general, only partial information on this motion is available in atmospheric models. For example the standard formulations of the Coriolis terms in atmospheric models take into account only the rotation of the Earth around its axis. Although it is generally accepted that the revolution of Earth and Moon does not induce Coriolis forces it is not obvious that the motion around the sun has no effect as well. These points can be clarified by deriving the complete guiding velocity.

It is common in atmospheric dynamics to assume a balance of centrifugal forces and the gravitational attractions by Sun and Moon. This implies that the gradients of centrifugal and gravitational potential cancel. The impact on the equatorial bulge is also taken into account. We are, however, not aware that this balance has been established in detail. It will be shown below that some aspects of this problem have not been clarified so far.

A further intriguing aspect of the Earth's motion will also be discussed briefly. What happens if the revolution is replaced by rotation? This must have an effect on the Coriolis and other inertial forces.

All this suggests to turn to a set of flow equations where these questions can be addressed properly. The framework of covariant flow equations provides an exact formulation of the guiding velocity, the Coriolis forces and the centrifugal terms and offers a possibility to deal with such problems in a coherent fashion. The formulation of covariant flow equations is a standard technique (e.g. Aris 1962) but we are not aware that the effect of the Earth's motion around the Sun and the Moon have been investigated this way.

It is the purpose of this work to derive the corresponding terms in the atmospheric flow equations and to compare them to standard formulations.

# 2. Equations

The Earth-Moon-Sun system is depicted in Fig. 1. The Moon's orbital plane is inclined by 5 deg with respect to the ecliptic plane. This angle is so small that it can be assumed to vanish. The related angular velocity vector $\mathbf{\Omega}_1$ is then perpendicular to the ecliptic plane and has a period of 27 days. The centers of Earth and Moon rotate around their barycenter (bullet). The vector $\mathbf{r}_1$ of length of $a_1 = 4.6 \cdot 10^6$ m connects the center of the Earth with this barycenter. The Earth rotates around the sun with a period of one year, an angular velocity $\Omega_2 = 2\pi/\text{year}$ and a radius $a_2 = 1.5 \cdot 10^{11}$ m. The Earth's rotation axis with angular velocity $\Omega_0 = 2\pi/\text{day}$ is inclined by the angle $\alpha = 27$ deg to the ecliptic plane. The Earth's radius is $a_0 = 6.4 \cdot 10^6$ m.

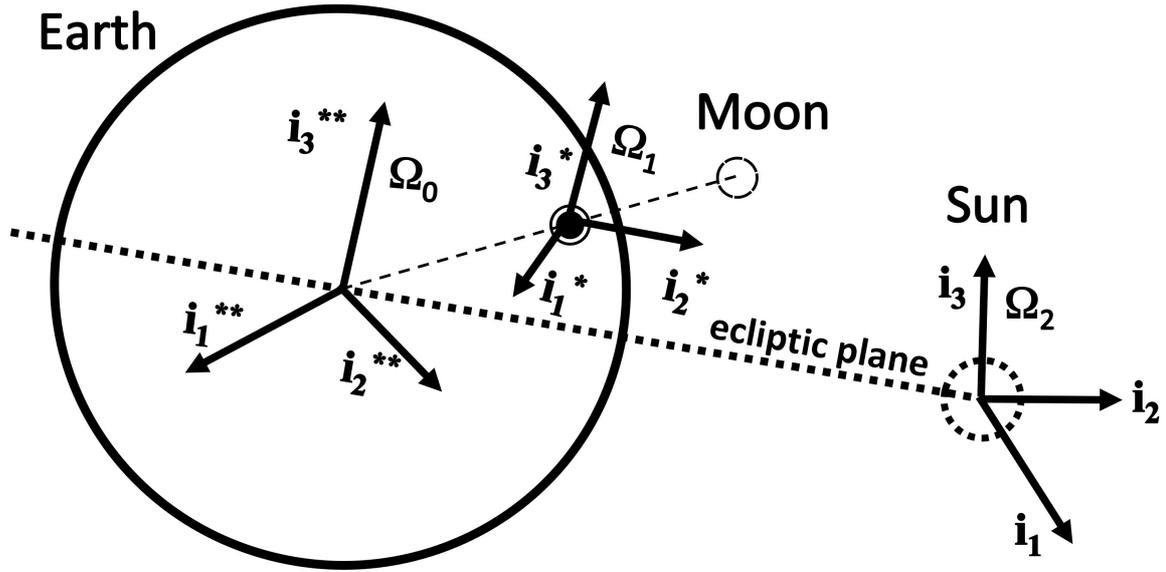

**Figure 1**: Sketch of the rotating Earth-Moon-Sun system: The vectors $\mathbf{i}_i$, $\mathbf{i}_i^*$, and $\mathbf{i}_i^{**}$ are the basic unit vectors of the inertial coordinates of Sun, Earth-Moon and Earth, respectively. The center of masses of the Earth-Moon system (barycenter) is indicated by the circled bullet, with a distance from the center of the Earth $a_1 = 4.8 \cdot 10^6$ m, which rotates with the Moon's angular velocity of $\Omega_1 \sim 27$ days (around the circled bullet); the related angular velocity vector $\mathbf{\Omega}_1$ is almost perpendicular to the ecliptic plane. The Earth's angular velocity axis $\mathbf{\Omega}_0$ has an $\alpha \sim 24$ deg inclination to the ecliptic plane and a rotation period of 1 day. The Sun's angular velocity axis $\mathbf{\Omega}_2$ has a rotation period of 1 year.

The formulation of the equations of motion must be based on a Galilean absolute system which is fixed with respect to the stars. The sun is chosen as origin (see Fig. 1) with basic unit vectors $\mathbf{i}_i$ ($i = 1 - 3$). These vectors are oriented such that $\mathbf{i}_3$ is aligned with $\mathbf{\Omega}_2$, while $\mathbf{i}_1$ and $\mathbf{i}_2$ are embedded in the ecliptic plane. The coordinates are ($x^1 = x$, $x^2 = y$, $x^3 = z$) so that the position vector is

$$\mathbf{r} = x\,\mathbf{i}_1 + y\,\mathbf{i}_2 + z\,\mathbf{i}_3 = x^i\,\mathbf{i}_i \qquad (1)$$

with summation convention. The barycenter is chosen as the origin of a further system with orthogonal basic vectors $\mathbf{i}_i^*$ where $\mathbf{i}_3^* = \mathbf{i}_3$ and coordinates $(x^*, y^*, z^*)$. Thus

$$\mathbf{i}_1^* = \mathbf{i}_1 \cos(\Omega_2 t) - \mathbf{i}_2 \sin(\Omega_2 t)$$
$$\mathbf{i}_2^* = \mathbf{i}_1 \sin(\Omega_2 t) + \mathbf{i}_2 \cos(\Omega_2 t). \qquad (2)$$

The third system with vectors $\mathbf{i}_i^{**}$ is that of the Earth where the 'vertical' vector

$$\mathbf{i}_3^{**} = \mathbf{i}_2 \sin(\alpha) + \mathbf{i}_3 \cos(\alpha) \qquad (3)$$

is aligned with the angular velocity of the Earth and does not depend on time. Let us mainly discuss the realistic case of revolution where the vectors $\mathbf{i}_1^{**}$ and $\mathbf{i}_2^{**}$ do not rotate around the barycenter. It is convenient for the formulation of the flow equation to introduce also the spherical coordinates $q^1 = \lambda$ for longitude, $q^2 = \varphi$ for latitude and $q^3 = a_0 + z = r$ for the distance to the Earth's center so that

$$x^{**} = r \cos(\varphi) \cos(\lambda)$$
$$y^{**} = r \cos(\varphi) \sin(\lambda) \qquad (4)$$
$$z^{**} = r \sin(\varphi).$$

.

The covariant atmospheric flow equations are presented, for example, in Kahlig (1974). The formulation of these equations for the Earth-Moon-Sun problem appears to have not been performed yet. This derivation is straightforward provided the functional dependence of the co-ordinates $x^i$ on the coordinates $q^i$ is known. The covariant basis vectors $\mathbf{q}_1 = \partial \mathbf{r}/\partial \lambda$, $\mathbf{q}_2 = \partial \mathbf{r}/\partial \varphi$, $\mathbf{q}_3 = \partial \mathbf{r}/\partial z$ are thus available. They are tangential to the latitude circles, the meridians and normal to the Earth's surface.

The Earth's axis is tilted and $\mathbf{i}_3^{**} = \mathbf{i}_2^* \sin(\alpha) + \mathbf{i}_3^* \cos(\alpha)$ does not depend on time. The basic vector

$$\mathbf{i}_2^{**} = \epsilon_3 \mathbf{i}_3^* + \epsilon_2 \mathbf{i}_2^* \qquad (5)$$

points towards the Sun and does not depend on $\Omega_1$. The coefficients $\epsilon_2$ and $\epsilon_3$ follow from the condition $\mathbf{i}_3^{**} \cdot \mathbf{i}_2^{**} = 0$ and $\mathbf{i}_2^2 = 1$ so that with (5)

$$\epsilon_3 \cos(\alpha) + \epsilon_2 \cos(\Omega_2 t) \sin(\alpha) = 0 \qquad (6)$$

$$\epsilon_2^2 + \epsilon_3^2 = 1.$$

Thus

$$\epsilon_2^2 = \cos^2(\alpha) \{ \cos^2(\Omega_2 t) \sin^2(\alpha) + \cos^2(\alpha) \}^{-1} \qquad (7)$$

$$\epsilon_3^2 = \epsilon_2^2 \cos^2(\Omega_2 t) \tan^2(\alpha)$$

and $\epsilon_3 = \sin(\alpha)$, $\epsilon_2 = \cos(\alpha)$ for $\Omega_2 = 0$. The third vector

$$\mathbf{i}_1^{**} = \gamma_1 \mathbf{i}_1 + \gamma_2 \mathbf{i}_2 + \gamma_3 \mathbf{i}_3 \qquad (8)$$

has to be orthogonal to $\mathbf{i}_2^{**}$ and $\mathbf{i}_3^{**}$. This leads to

$$\gamma_3^2 = \sin^2(\alpha) \{ (\epsilon_2 \cos(\Omega_2 t) \cos(\alpha) - \epsilon_3 \sin(\alpha))^2 / (\epsilon_2 \sin(\Omega_2 t))^2 + 1 \}^{-1}$$

$$\gamma_1^2 = ( 1 - \gamma_3^2 ( \cotan^2(\alpha) + 1) )^{-1} \qquad (9)$$

$$\gamma_2 \sin(\alpha) + \gamma_3 \cos(\alpha) = 0$$

$$\gamma_3 \epsilon_3 + \gamma_1 \epsilon_2 \sin(\Omega_2 t) + \gamma_2 \epsilon_2 \cos(\Omega_2 t) = 0.$$

Note that for $\gamma_2 = \gamma_3 = 0$, $\gamma_1 = 1$ for $\Omega_2 = 0$. All $\epsilon_i$ and $\gamma_i$ depend on time if $\Omega_2 = 0$. With that we can equate the position vector $\mathbf{r}$ (see (1)) to the position vector

$$\mathbf{r}^{**} = \mathrm{x}^{**} \mathbf{i}_1^{**} + \mathrm{y}^{**} \mathbf{i}_2^{**} + \mathrm{z}^{**} \mathbf{i}_3^{**} \qquad (10)$$
$$- (a_2 - a_1) ( \mathbf{i}_1 \sin(\Omega_2 t) + \mathbf{i}_2 \cos(\Omega_2 t) ) + a_1 ( \mathbf{i}_1^* \sin(\Omega_1 t) - \mathbf{i}_2^* \cos(\Omega_1 t) ),$$

which contains the vectors pointing from the sun to the barycenter and from that to the Earth. The result is the relation of all $x^i$ to all $q^i$, namely,

$$x = a_2 \sin(\Omega_2 t) - a_1 \sin(\Omega_1 t) + r \cos(\varphi) \{ \gamma_1 \cos(\lambda + \Omega_0 t) + \epsilon_2 \sin(\lambda + \Omega_o t) \sin(\Omega_2 t) \}$$

$$y = a_2 \cos(\Omega_2 t) + a_1 \cos(\Omega_1 t) + r \cos(\varphi)\{\gamma_2 \cos(\lambda + \Omega_0 t) + \epsilon_2 ( \sin(\lambda + \Omega_0 t) \cos(\Omega_2 t) \}$$

$$+ r \sin(\varphi) \sin(\alpha) \qquad (11)$$

$$z = r \cos(\varphi) \{ \gamma_3 \cos(\lambda + \Omega_0 t) + \epsilon_3 \sin(\lambda + \Omega_0 t) \} + r \sin(\varphi) \cos(\alpha).$$

All $\mathbf{q}_j$ follow immediately from (11). The metric of this system is given by $g_{ij} = \mathbf{q}_i \cdot \mathbf{q}_j$ with

$$g_{11} = r^2 \cos^2(\varphi), \ g_{22} = r^2, \ g_{33} = 1 \qquad (12)$$

as in standard spherical coordinates. The absolute velocity is, of course,

$$\mathbf{v}_A = \mathbf{i}_1 \, dx/dt + \mathbf{i}_1 \, dy/dt + \mathbf{i}_1 \, dz/dt, \qquad (13)$$

but this velocity must be expressed in terms of the covariant basis vectors. Thus

$$\mathbf{v}_A = \partial \mathbf{r}/\partial t|_{qi} + \partial \mathbf{r}/\partial q^i \, dq^i/dt = \partial \mathbf{r}/\partial t + \mathbf{q}_i \, dq^i/dt. \qquad (14)$$

The 'guiding velocity'

$$\mathbf{W} = \mathbf{v}_A - \mathbf{q}_i \, dq^i/dt - \partial \mathbf{r}/\partial t \qquad (15)$$

is the difference of absolute velocity and relative velocity

$$\mathbf{v}_R = \mathbf{q}_i \, dq^i/dt, \qquad (16)$$

which describes the atmospheric motions. The covariant flow equations are based on (11) - (16). The specific form of equations used here relies on the unpublished lecture notes 'Dynamische Gleichungen in allgemeinen Koordinaten' by the late Prof. Hinkelmann, which have been selected because of their convenient notation. The equations are

$$dv_k/dt - \tfrac{1}{2} v^n v^m \, \partial g_{nm}/\partial q_k + 2\omega_{nk} v^n = -(1/\rho) \nabla_k p - \nabla_k(\Phi_a - \tfrac{1}{2} P_C) - \partial W_k/\partial t \qquad (17)$$

where friction terms are deleted. The $q^k$ (k = 1, …,3) in (17) are general coordinates with $q^1$ = λ, $q^2$ = φ, and $q^3$ = r in our case. The derivatives in the operator $\nabla_k$ are to be carried out with respect to these coordinates. The relative velocity components are $v_k = g_{kn}\, dq^n/dt$ and $v^k = dq^k/dt$. Those of **W** are

$$W_k = \partial x^i/\partial q_k\ \partial x^i/\partial t. \tag{18}$$

The last lhs term in (17) is the Coriolis term with

$$\omega_{ij} = -\tfrac{1}{2}\,(\partial W_i/\partial q^j - \partial W_j/\partial q^i). \tag{19}$$

The potential of the external forces is $\Phi_a$, that of the centrifugal forces is $\tfrac{1}{2}P_C$ with

$$P_C = (\partial x_i/\partial t)^2. \tag{20}$$

The first lhs term in (17) is the total derivative of the relative velocity $v_k$, the second term is sometimes called "curvature term" (Holton, 1992). These terms are well known from the standard forms of spherical flow equations except that it is common practice to predict the velocities $u = g_{11}^{1/2}\, d\lambda/dt$, $v = g_{22}^{1/2}\, d\varphi/dt$, $w = dr/dt$. A transformation of (17) to this standard form is achieved by dividing all terms of (17) by $g_{kk}^{1/2}$ and adding a further curvature term for k = 1. The formulation of these terms is little affected by the switch to the covariant formulation and will, therefore, not be further commentated on. The Coriolis term (19) involves lengthy calculations for a complicated guiding velocity as in our case. The results of these evaluations will be discussed below. The pressure gradient term is of standard form. The gradients of the gravity potential and of the centrifugal potential follow. The last rhs term is the local time derivative of the components of the guiding velocity. It is common practice in atmospheric models to assume a cancellation of the horizontal gradients of $\Phi_a$ and $P_C$ so that the second rhs term in (17) vanishes for k = 1 and k = 2. That implies that the last term in (17) is omitted as well.

## 3. Guiding velocity

Inserting (11) in (18) we obtain the components $W_k$ of the guiding velocity. One finds

$$W_1 = r^2 \cos^2(\varphi)\{ \Omega_0 + \Omega_1 \, a_1 \, (r \cos(\varphi) \, (\sin(\lambda + \Omega_0 t) \cos(\Omega_1 t)$$
$$- \cos(\lambda + \Omega_0 t) \sin(\Omega_1 t) \cos(\alpha) )\} \quad (21)$$

Division by $g_{11}^{1/2} = r \cos(\varphi)$ is required to obtain velocities of dimension (m s$^{-1}$). The first rhs term represents the rotation of the Earth while the second term stems from the revolution. The contribution by this term is small with $\Omega_1 \, a_1 / (\Omega_0 \, a_0) \sim (1/27)(4.7/6.6) = 2.5 \times 10^{-2}$ when compared to the first term but this term depends on the longitude and the velocity $\Omega_1 \, a_1 \sim$ 360 m s$^{-1}$ is large. There are further frequencies $\Omega_0 \pm \Omega_1$. The formula for $W_1$ becomes rather involved if the rotation around the sun is included. The result is

$$W_1 = r^2 \cos^2(\varphi) \{ \Omega_0 - \Omega_2 \, (\Gamma_1 \, \epsilon_2 \sin(\Omega_2 t) \quad (22)$$
$$- E_2 \, \gamma_2 \cos(\Omega_2 t) + \Gamma_3 \, \epsilon_3 \cos^2(\lambda + \Omega_0 t) - E_3 \, \gamma_3 \sin^2(\lambda + \Omega_0 t) \}$$
$$- \sin^2(\lambda + \Omega_0 t) \, d/dt \, \{ \epsilon_2 \, \gamma_1 \sin(\Omega_2 t) \} - \cos^2(\lambda + \Omega_0 t) \, d/dt \{ \epsilon_2 \, \gamma_2 \cos(\Omega_2 t) \}$$
$$+ \partial x / \partial \lambda \, \{ a_2 \, \Omega_2 \cos(\Omega_2 t) + a_1 \, \Omega_1 \cos(\Omega_1 t) \}$$
$$+ \partial y / \partial \lambda \, \{ a_2 \, \Omega_2 \sin(\Omega_2 t) + a_1 \, \Omega_1 \sin(\Omega_1 t) \}$$

where

$$\Omega_2 \, \Gamma_i = d\gamma_i / dt \quad (23)$$
$$\Omega_2 \, E_i = d\epsilon_i / dt.$$

The time derivatives act on functions of $\Omega_2 \, t$ so that the related terms are $\sim \Omega_2$ just as the second rhs term. Division of $W_1$ by $g_{11}^{1/2} = r \cos(\varphi)$ yields velocities. The first rhs term $\sim \Omega_0$ describes the rotation of the Earth, but is also affected by the complicated relation of the Earth's basic vectors and those of the sun. Nevertheless, the latter ones are of the order of $a_0 \, \Omega_2 \sim 1$ ms$^{-1}$. They are presumably not negligible but have not been discussed so far in the literature. They contain contributions with a semi-diurnal component and also annual oscillation but one would need a detailed frequency analysis of the factors $\gamma_1 \, \epsilon_1, \gamma_2 \, \epsilon_2$ etc. to determine all frequencies exactly. The last two terms are related to the motion of the Earth's center. They are extremely large with $a_2 \Omega_2 \sim 2 \cdot 10^6$ ms$^{-1}$.

The meridional component $W_2$ is as complicated as the zonal one with

$$W_2 = \Omega_0 r^2 \sin(\varphi) \cos(\varphi) \{ \gamma_3 \epsilon_2 (\cos^2(\lambda + \Omega_0 t) - \sin^2(\lambda + \Omega_0 t)) + \gamma_2 \epsilon_2 \cos(\Omega_2 t)$$
$$- \gamma_3 \epsilon_2 \cos(\Omega_2 t) \sin(\Omega_2 t) \}$$
$$- \Omega_2 r^2 \sin(\varphi) \cos(\varphi) \cos(\lambda + \Omega_0 t)) \sin(\lambda + \Omega_0 t) (\Gamma_1 \epsilon_2 + \gamma_1 E_2 + \Gamma_2 \epsilon_2 - \gamma_2 \epsilon_2)$$
$$- r \sin(\alpha) \cos(\varphi) \partial y/\partial t + r \cos(\alpha) \cos(\varphi) \partial z/\partial t \qquad (24)$$
$$+ \{ a_2 \Omega_2 \cos(\Omega_2 t) - a_1 \Omega_1 \cos(\Omega_1 t) \} \partial x/\partial \varphi$$
$$- \{ a_2 \Omega_2 \sin(\Omega_2 t) + a_1 \Omega_1 \sin(\Omega_1 t) \} \partial y/\partial \varphi.$$

The first term $\sim \Omega_0$ is quite large. There are contributions due to the tilt and terms related to the motion of the Earth's center as in (22). The components of **W** are needed to derive the explicit formulation of the Coriolis terms.

## 4. Coriolis term

One may expect a rather complicated form of the Coriolis terms in view of (22). However, it simplifies the formulas that the Earth-Moon revolution does not induce Coriolis terms. Inserting (17) into (18) yields

$$2 \omega_{i12} = \partial x/\partial \varphi \, \partial^2 x/\partial \lambda \partial t - \partial x/\partial \lambda \, \partial^2 x/\partial \varphi \partial t + \partial y/\partial \varphi \, \partial^2 y/\partial \lambda \partial t - \partial y/\partial \lambda \, \partial^2 y/\partial \varphi \partial t$$
$$+ \partial z/\partial \varphi \, \partial^2 z/\partial \lambda \partial t - \partial z/\partial \lambda \, \partial^2 z/\partial \varphi \partial t. \qquad (25)$$

Using (9) and (11) one finds

$$2 \omega_{i12} = 2\Omega_0 r^2 \sin(\varphi) \cos(\varphi) + r^2 \sin(\varphi) \cos(\varphi) \Omega_2 [ (\gamma_2 E_2 - \Gamma_2 \epsilon_2 + \gamma_1 \epsilon_2) \cos(\Omega_2 t)$$
$$- (-\Gamma_1 \epsilon_2 + \gamma_1 E_2 - \gamma_2 \epsilon_2) \sin(\Omega_2 t) + \gamma_3 E_3 - \Gamma_3 \epsilon_3$$
$$+ \sin(\alpha) \{ \epsilon_2 \cos((\lambda + \Omega_2 t) \sin(\Omega_2 t) + \Gamma_2 \sin(\lambda + \Omega_0 t) - E_2 \sin(\lambda + \Omega_0 t) \}$$
$$+ \cos(\alpha) \{ \Gamma_3 \sin(\lambda + \Omega_0 t) - E_3 \cos(\lambda + \Omega_0 t) \} ] \qquad (26)$$

The first rhs term, when multiplied by $d\varphi/dt$ and divided by $r\cos(\varphi)$, turns into the standard form of the Coriolis term. All other terms are not standard and relatively small because of the faster $\Omega_2$. They have about the same order of magnitude as the Coriolis term $2w\,\Omega_0\cos(\varphi)$ in the first equation of motion. A gross estimate W for the vertical velocity follows from $W \sim HU/L$ with horizontal velocity U, height and length scale H and L. Thus $W\,\Omega_0 \sim U\,\Omega_2$, if $H/L \sim 1/365$, that is, for a reasonable relation of H/L. The new Coriolis terms in (26) depend on time with nearly diurnal and annual frequencies.

The vertical Coriolis term in the first equation of motion is

$$2\,\omega_{13} = 2\Omega_0\,r^2\cos^2(\varphi) + \qquad (27)$$

$$+ r^2\cos^2(\varphi)\,\Omega_2\,\{\gamma_2\,\epsilon_2\sin(\Omega_0 t) - \gamma_1\,\epsilon_2\cos(\Omega_2 t) + (\Gamma_1\,\epsilon_2 - \gamma_1\,E_2)\sin(\Omega_2 t)$$

$$+ (\Gamma_2\,\epsilon_2 - \gamma_2\,E_2)\cos(\Omega_2 t) + \Gamma_3\,\epsilon_3 - \gamma_3\,E_3\,\}$$

$$+ r\sin(\varphi)\cos(\varphi)\,\Omega_2\,[\,\sin(\alpha)\,\{\,\Gamma_2\sin(\lambda + \Omega_0 t)\sin(\Omega_2 t) - E_2\cos(\Omega_2 t)\,\}$$

$$- \cos(\alpha)\,\{\,\Gamma_3\sin(\lambda + \Omega_0 t) + E_3\cos(\lambda + \Omega_0 t)\,\}\,].$$

The first rhs term is the standard Coriolis term, the others are all $\sim \Omega_2$. They would be of interest only in a nonhydrostatic model. The tilt of the Earth's induces terms of the same order of magnitude. Thus the rotation around the Sun induces Coriolis terms with a diurnal and an annual cycle. Moreover, terms $\sim \cos(\varphi)$ do not vanish at the equator and are thus of particular dynamic interest. The frequency $\Omega_1$ is not found in (26) and (27).

## 5. Centrifugal accelerations

The centrifugal acceleration in (17) is given by the term $F_{Ck} = -\frac{1}{2}\nabla_k p - \partial W_k/\partial t$. It is not immediately obvious that $\mathbf{F}_C$ is the gradient of a potential $\Pi$. To show this one would have to cast the tendency term in the form of a gradient. Combining (11) with (18) and (20) we obtain

$$F_{Ck} = \partial/\partial q^k\,(\tfrac{1}{2}\,(\partial x^i/\partial t)^2) - \partial/\partial t\,(\partial x^i/\partial q^k\,\partial x^i/\partial t) = -\partial^2 x^i/\partial t^2\,\partial x^i/\partial q^k \qquad (28)$$

for the components of **F**$_C$. The contribution of P$_C$ to F$_C$ drops out in (28). This does, of course, not mean that P$_C$ is unimportant but (28) stresses the dominant role of the tendency term. The second derivative with respect to time as in (28) can only stem from $\partial W_k/\partial t$. This point appears to have received little attention so far.

If there exists a potential $\Pi$, it has to satisfy the equations

$$\partial^2 x^i/\partial t^2 \; \partial x^i/\partial q^k = \partial \Pi/\partial q^k \qquad (29)$$

and, therefore,

$$\partial/\partial\varphi \; (\partial x_i/\partial t)^2 \; \partial x/\partial\lambda = \partial/\partial\lambda \; (\partial x_i/\partial t)^2 \; \partial x_i/\partial\varphi$$

without summation convention in the factor $(\partial x_i/\partial t)^2$. Inserting (11) and (28) yields

$$\begin{aligned}
F_{C1} \; &= 2\Omega_0 \, r^2 \cos^2(\varphi) \, \{ \, d/dt \, (\gamma_1 \, \epsilon_1 \sin(\Omega_2 t) + \gamma_1 \, \epsilon_1 \cos(\Omega_2 t) \qquad (30)\\
&+ (-\, a_2 \, \Omega_2^2 \sin(\Omega_2 t) + a_1 \, \Omega_1^2 \sin(\Omega_1 t)) \, \partial x/\partial\lambda \\
&- (a_2 \, \Omega_2^2 \sin(\Omega_2 t) - a_1 \, \Omega_1^2 \sin(\Omega_1 t)) \, \partial y/\partial\lambda \, \}
\end{aligned}$$

$$\begin{aligned}
F_{C2} \; &= \Omega_0 \, r^2 \sin(\varphi) \cos(\varphi) + \qquad (31)\\
&+ 2\Omega_0 \, \Omega_2 \, r^2 \sin(\varphi) \cos(\varphi) \, [\, -(\gamma_1 \, E_2 + \gamma_1 \, \epsilon_2) \sin(\lambda + \Omega_0 t) \cos(\lambda + \Omega_0 t) \cos(\Omega_2 t) \\
&- \Gamma_1 \, \epsilon_2 \sin(\lambda + \Omega_0 t) + \{\, \Gamma_2 \, \epsilon_2 \sin^2(\lambda + \Omega_0 t) - E_2 \, \gamma_2 \cos^2(\lambda + \Omega_0 t) \,\} \cos(\Omega_2 t) \\
&+ \epsilon_2 \, \gamma_2 \cos^2(\lambda + \Omega_0 t) \} \sin(\Omega_2 t) + \epsilon_3 \, \Gamma_3 \sin^2(\lambda + \Omega_0 t) - E_3 \, \gamma_3 \cos^2(\lambda + \Omega_0 t) \\
&+ 2\Omega_0 \, \Omega_2 \, r^2 \cos^2(\varphi) \sin(\Omega_2 t) \sin(\alpha) - \{a_2 \, \Omega_2^2 \cos(\Omega_2 t) - a_1 \Omega_1^2 \cos(\Omega_1 t)\} \, \partial y/\partial\varphi \\
&- \{\, a_2 \, \Omega_2^2 \sin(\Omega_2 t) - a_1 \Omega_1^2 \sin(\Omega_1 t) \,\} \, \partial x/\partial\varphi
\end{aligned}$$

The terms $\sim \Omega_2^2$ are omitted except in combination with $a_2$. There is no term $\sim \Omega_0^2$ in (28) and the time derivatives are equivalent to factors $\Omega_2$. The second and third terms in (28) are clearly centrifugal. They are quite large with $a_2 \, \Omega_2^2 \sim 6 \; 10^{-3} \text{ ms}^{-2}$ while $\Omega_0 \, \Omega_2 \, a_2 \sim 6 \; 10^{-5} \text{ ms}^{-2}$. However, the centrifugal terms are balanced by gravitational attraction as will be demonstrated below. Thus, the first term in (28) is not negligible.

The first rhs term in (29) is well known and represents the centrifugal acceleration due to the Earth's rotation. This term is balanced by the influence of the equatorial bulge and, therefore, not included in atmospheric models. There is a lengthy expression $\sim \Omega_0 \Omega_2$ and centrifugal contributions due to the motion of the Earth's center.

The acceleration $F_{C3}$ is not given but may be of interest in a nonhydrostatic model. The condition (29) has been checked but no violation has been found.

## 6. Gravitational attraction

The forces of gravity vary $\sim 1/d^2$ with distance d between the respective centers of mass. Thus, the solar attraction can be assumed constant in the atmosphere because $a_0/a_2$ is so small. Variation of attraction by the moon is not negligible and is taken into account in tidal theories (e.g. Henderschott 2005). The gravity forces are expected to almost compensate the centrifugal forcings $F_{C1}$ and $F_{C2}$. Let us concentrate on the solar impact. The unit vector $\mathbf{S}_2 = \mathbf{i}_1 \sin(\Omega_2 t) + \mathbf{i}_2 \cos(\Omega_2 t)$ points to the Sun. The basic vectors $\mathbf{i}_i$ must be expressed in terms of contravariant vectors in order to insert the acceleration into the equation of motion. The $\mathbf{q}^i$ are related by

$$\mathbf{q}^i = 1/(g_{ii})^{½} \, \mathbf{q}_i$$

to the covariant basic vectors in our case and few calculations are needed to show that

$$\mathbf{i}_1 = (r\cos(\varphi))^{-1} \, \partial x/\partial \lambda \, \mathbf{q}^1 + (1/r) \, \partial x/\partial \varphi \, \mathbf{q}^2 + \partial x/\partial r \, \mathbf{q}^3 \qquad (32)$$
$$\mathbf{i}_2 = (r\cos(\varphi))^{-1} \, \partial y/\partial \lambda \, \mathbf{q}^1 + (1/r) \, \partial y/\partial \varphi \, \mathbf{q}^2 + \partial y/\partial r \, \mathbf{q}^3$$

The vector $a_2 \Omega_2^2 \mathbf{S}_2$ represents the gravitational attraction by the sun assuming as usual an exact balance of attraction and centrifugal force in the center of the Earth. Thus $a_2 \Omega_2^2$ (r cos($\varphi$))$^{-1}$ $\partial x/\partial \lambda$ + (1/r) $\partial y/\partial \varphi$ has to balance the last two rhs terms of $F_{C1}$ (see (30)). This balance is exact. This can be shown also for $F_{C2}$. However, the first rhs term of $F_{C1}$ and all rhs terms $\sim \Omega_0 \Omega_2$ of $F_{C2}$ are not balanced and should be included in the atmospheric models. The related accelerations are not negligible with $a_0 \Omega_0 \Omega_2 \sim 10^{-4}$ ms$^{-2}$.

As mentioned above, the situation with respect to the moon is more difficult to handle, because the lunar attraction decreases by a factor of $2a_0/a_m \sim 1.6 \cdot 10^{-5}$ ms$^{-2}$. The corresponding formulas have been derived but not presented, because we can rely here on standard tidal theory. Nevertheless, the corresponding tidal effect is not negligible.

## 7. Rotation instead of revolution

As mentioned, it is interesting to have a look at the situation when Earth and Moon rotate around each other. The basic vectors $\mathbf{i}_i^*$ must be replaced by $\hat{\mathbf{i}}_i^*$ where then $\hat{\mathbf{i}}_1^* = \mathbf{i}_1^*$ and $\hat{\mathbf{i}}_2^* = \mathbf{i}_2$. Moreover, $\hat{\mathbf{i}}_3^* \mathbf{x}^* = \mathbf{i}_3^{**}$, but

$$\hat{u}_1^{**} = u_1 \cos((\Omega_1 + \Omega_2)t) + u_2 \cos(\alpha) \sin((\Omega_1 + \Omega_2)t) - u_3 \sin(\alpha) \sin((\Omega_1 + \Omega_2)t) \quad (33)$$

$$\hat{u}_2^{**} = - u_1 \sin((\Omega_1 + \Omega_2)t) + u_2 \cos(\alpha) \cos((\Omega_1 + \Omega_2)t) - u_3 \sin(\alpha) \cos((\Omega_1 + \Omega_2)t)$$

so that there is now a simple addition of both angular velocities. The relation of the coordinates in the case of rotation is

$$\hat{x} = a_2 \sin(\Omega_2 t) - a_1 \sin((\Omega_1 - \Omega_2)t) - r \cos(\varphi) \cos(\lambda + \hat{\Omega} t)$$

(34)

$$\hat{y} = a_2 \cos(\Omega_2 t) + a_1 \cos((\Omega_1 - \Omega_2)t) - r \cos(\varphi) \cos(\alpha) \sin(\lambda + \hat{\Omega} t) - r \sin(\varphi) \sin(\alpha)$$

$$\hat{z} = - r \cos(\varphi) \sin(\alpha) \sin(\lambda + \hat{\Omega} t) + r \sin(\varphi) \cos(\alpha)$$

where $\hat{\Omega} = \Omega_1 + \Omega_2 + \Omega_3$ is close to $\Omega_0$. The calculations of the Coriolis terms use, of course, the same formalism as above and yield

$$2 \omega_{12} = 2 \hat{\Omega} \sin(\varphi), \quad (35)$$

so that the inclusion of Moon and Sun lead to a small modification of the standard result. The accelerations $F_{Ci}$ have been evaluated as well. For example,

$$\hat{F}_{C1} = a_2 \Omega_0 r \cos(\varphi) \{ \sin(\Omega_2 t) + \gamma_1 \epsilon_1 \cos(\Omega_2 t) \quad (36)$$

$$+ (- a_2 \Omega_2^2 \sin(\Omega_2 t) + a_1 \Omega_1^2 \sin(\Omega_1 t)) \, \partial x/\partial \lambda$$

$$- (a_2 \Omega_2^2 \sin(\Omega_2 t) - a_1 \Omega_1^2 \sin(\Omega_1 t)) \, \partial y/\partial \lambda \, \}.$$

Thus, only the centrifugal terms are left and there are no additional unbalanced terms as in the case of revolution. It follows that almost all complications listed in sections 3 and 4 are due to revolution.

## 8. Conclusions

This work has been stimulated by the notion that atmospheric global models do not take into account the Earth-Moon revolution nor the rotation around the sun except for the compensation of gravitational attraction and centrifugal forces in the center of the Earth. The covariant form of the equations of motion has been chosen to deal with these uncertainties because they contain a complete form of the Coriolis terms and the centrifugal acceleration **Fc**. This formalism can be applied if the relation between the coordinates of the resting absolute solar system and the standard spherical coordinates for the rotating Earth is available. This relation (11) appears not to have been used in atmospheric models as yet.

The Coriolis terms have been derived and discussed in detail for the first equation of motion. It is a surprising result that the rotation around the Sun induces Coriolis forces with frequencies close to $\Omega_0$ and $\Omega_2$, which are not negligible at least in nonhydrostatic models. As expected the revolution has no impact on Coriolis forces.

There is a clear formulation for all centrifugal accelerations which are calculated explicitly. It turns out that the local time derivative of the guiding velocity is a main contributor to these accelerations. It is shown that the compensation of centrifugal accelerations and gravitational attraction is quite effective for many of the centrifugal terms but there remain unbalanced accelerations which are not taken as yet into account in atmospheric models. It has been demonstrated that there exists a potential for the $F_{Ci}$ despite the complicated form of the guiding velocity. Replacing the revolution by rotation leads to a simple form of the Coriolis terms and also of the centrifugal forcing terms.

A few shortcomings of our approach should be noted. The motions of the Earth around the barycenter and the Sun have been assumed circular. Corresponding corrections are possible, of course, but it has been felt that the related calculations can be postponed at this early stage of investigations. Tests with advanced global models are needed to assess the relative importance of the dynamical modifications due to the Earth's motion in the solar system as derived here.

Any search for observational evidence for the impact of the new terms derived here is hampered by their smallness. It would be difficult to find support from data for the new Coriolis terms in (26) and (27) or for the new accelerations in (30) and (31). It may be possible, however, to find some support from tidal effects in the atmosphere due to the variation of lunar attraction. There is the obvious problem that the contribution of diurnal motions to atmospheric spectra is quite large compared to such tidal effects.

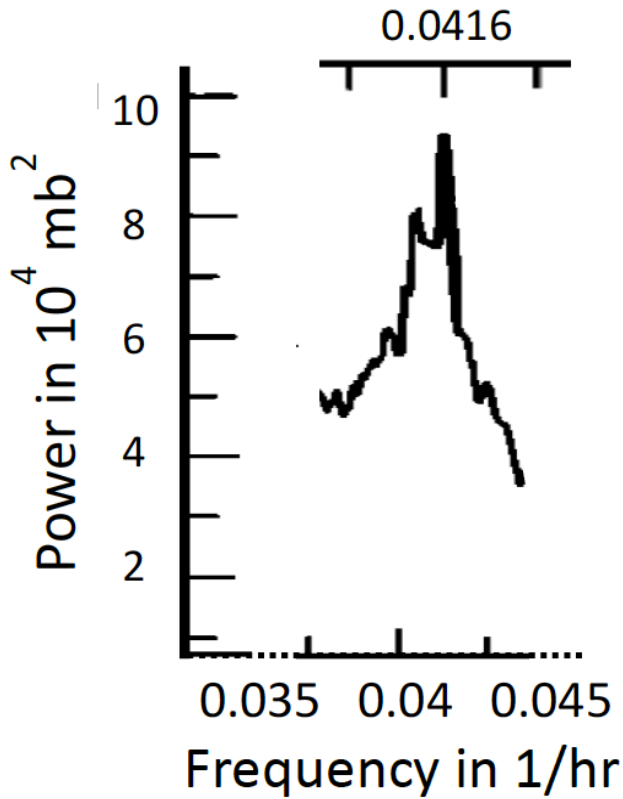

**Figure 2**: Power spectrum of hourly barometric data from Batavia (adapted from Figure 2 of Hamilton and Garcia, 1986) given as power per frequency interval of one inverse year (or 0.000114 h$^{-1}$). The dominating peak (solar diurnal tide) is located at 0.0416 (see upper axis), and its left neighbour corresponds to the period $T_L$ of about 25 hours of the rotating Earth-Moon system discussed here. Only that part of the spectrum close to the diurnal cycle is shown.

A separation of effects is difficult. A filtering of the diurnal signal in atmospheric data should allow to see at least some evidence of contributions at the frequency 0.040 h$^{-1}$ corresponding to the period of the Earth-Moon system $T_L$ of 25 hours. Hamilton and Garcia (1986) applied such a filter to hourly surface pressure observations at various stations in the belt 5°S - 20°N. A related section of the power spectrum near these frequencies is presented in Fig. 2. Although the diurnal signal is still dominant despite the filtering there is a clearly visible smaller peak at frequency 0.040 h$^{-1}$. This peak is seen at all stations analyzed by Hamilton and Garcia (1986). This suggests that we see here indications of the effects of the tidal accelerations. At least none of the atmospheric global normal modes has a period close to 25 hours (see Table 1 of Hamilton and Garcia, 1986).

Acknowledgement: Both authors are grateful to Dr. D. Lang for technical support; KF acknowledges support and scientific working environment provided by the Max Planck Institute of Meteorology.

References:

Andrews, D., J. Holton and C. Leovy, 1987: Middle Atmosphere Dynamics. Int. Geophys. Series 40, 489 pp.

Aris, R., 1962: Vectors, Tensors, and the Basic Equations of Fluid Mechanics, Dover Publications, New York, 286 pp.

Durran, D., 1993: Is the Coriolis force really responsible for the inertial oscillation? Bull. Amer. Meteor. Soc., 74, 2179–2184.

Egger, J., 2011: Mountain forces and the atmopheric energy budget. J. Atmos. Sci., 68, 2689-2694.

Hamilton, K. and R. R. Garcia, 1986: Theory and observations of the short-period normal mode oscillations of the atmosphere. J. Geophys. Res., 91, 11867-11875.

Hendershott, M., 2005: Introduction to ocean tides. 2004 Program of Study: Tides, Woods Hole Oceanog. Inst. Tech. Rept., WHOI-2005-08, 1-19.

Holton, J., 1992: An Introduction to Dynamic Meteorology. Acad. Press, San Diego, 511pp.

Kahlig, P., 1974: Über hydrodynamische Gleichungen bei zerlegter Metrik. Arc. Met. Geoph. Biokl. Ser. A, 23, 143-160.

Kahlig, P. 1974: Über die kinetischen Größen in sphärischen Koordinatensystemen mit zeitabhängiger Vertikalkoordinate. Met. Geoph. Biokl. Ser. A, 23, 55-64.

Kagan, B. and J. Sündermann, 1996: Dissipation of tidal energy, paleotide and the evolution of the earth-moon system. Adv. Geophys., 36, 179-275.

Lowrie, W., 2007: Fundamentals of Geophysics. Cambridge University Press, 381 pp.

Lyard, F., F. Lefevre, Th. Letellies and O. Francis, 2006: Modelling the global ocean tides: modern insights from FES2004. Ocean Dynamics, 56, 394-425.

Phillips, N. A., 2000: An explication of the Coriolis effect. Bull. Amer. Meteorol. Soc., 81, 299-303.

Vallis, G. K., 2006: Atmospheric and Oceanic Fluid Dynamics. Cambridge University Press, 745 pp.

White, A. and R. Bromley, 1995: Dynamically consistent, quasi-hydrostatic equations for global models with a complete representation of the Coriolis force. Quart. J. Roy. Met. Soc., 121, 399-418.